\documentclass[twocolumn,showpacs,prl,nofootinbib,superscriptaddress,amsmath,amssymb,floatfix]{revtex4}

\usepackage{graphicx}
\usepackage{dcolumn}
\usepackage{bm}

\newcommand{\gray}{$\gamma$-ray}

\newcommand{\hi}{H{\sc i}}
\newcommand{\htwo}{H$_{2}$}
\newcommand{\hii}{H{\sc ii}}

\newcommand{\Xco}{$X_{\rm CO}$}

\newcommand{\err}[2]{$#1 \pm #2$}
\newcommand{\scale}[1]{$\times \; 10^{-#1}$}

\newcommand{\gadget}{{\it GaDGET}{}}
\newcommand{\fermi}{{Fermi}{}}

\begin{document}

\preprint{}

\title{The Spectrum of the Isotropic Diffuse Gamma-Ray Emission Derived From 	First-Year \fermi{} Large Area Telescope Data}

\date{\today}

\author{A.~A.~Abdo}
\affiliation{Space Science Division, Naval Research Laboratory, Washington, DC 20375, USA}
\affiliation{National Research Council Research Associate, National Academy of Sciences, Washington, DC 20001, USA}
\author{M.~Ackermann}
\email{markusa@slac.stanford.edu}
\affiliation{W. W. Hansen Experimental Physics Laboratory, Kavli Institute for Particle Astrophysics and Cosmology, Department of Physics and SLAC National Accelerator Laboratory, Stanford University, Stanford, CA 94305, USA}
\author{M.~Ajello}
\affiliation{W. W. Hansen Experimental Physics Laboratory, Kavli Institute for Particle Astrophysics and Cosmology, Department of Physics and SLAC National Accelerator Laboratory, Stanford University, Stanford, CA 94305, USA}
\author{W.~B.~Atwood}
\affiliation{Santa Cruz Institute for Particle Physics, Department of Physics and Department of Astronomy and Astrophysics, University of California at Santa Cruz, Santa Cruz, CA 95064, USA}
\author{L.~Baldini}
\affiliation{Istituto Nazionale di Fisica Nucleare, Sezione di Pisa, I-56127 Pisa, Italy}
\author{J.~Ballet}
\affiliation{Laboratoire AIM, CEA-IRFU/CNRS/Universit\'e Paris Diderot, Service d'Astrophysique, CEA Saclay, 91191 Gif sur Yvette, France}
\author{G.~Barbiellini}
\affiliation{Istituto Nazionale di Fisica Nucleare, Sezione di Trieste, I-34127 Trieste, Italy}
\affiliation{Dipartimento di Fisica, Universit\`a di Trieste, I-34127 Trieste, Italy}
\author{D.~Bastieri}
\affiliation{Istituto Nazionale di Fisica Nucleare, Sezione di Padova, I-35131 Padova, Italy}
\affiliation{Dipartimento di Fisica ``G. Galilei", Universit\`a di Padova, I-35131 Padova, Italy}
\author{B.~M.~Baughman}
\affiliation{Department of Physics, Center for Cosmology and Astro-Particle Physics, The Ohio State University, Columbus, OH 43210, USA}
\author{K.~Bechtol}
\affiliation{W. W. Hansen Experimental Physics Laboratory, Kavli Institute for Particle Astrophysics and Cosmology, Department of Physics and SLAC National Accelerator Laboratory, Stanford University, Stanford, CA 94305, USA}
% ...

\begin{abstract}
We report on the first \fermi\
Large Area Telescope (LAT) measurements of the 
so-called ``extragalactic'' diffuse \gray{} emission (EGB).
This component of the diffuse \gray{} emission 
is generally considered to have an
isotropic or nearly isotropic distribution on the sky with diverse
contributions discussed in the literature.
The derivation of the EGB is based on 
detailed modelling of the bright foreground diffuse 
Galactic \gray{} emission (DGE), the detected 
LAT sources and the solar \gray{} emission.
We find the spectrum of the EGB is 
consistent with a power law with differential 
spectral index $\gamma = 2.41 \pm 0.05$ and intensity,  
$I(> 100\,{\rm MeV}) = (1.03 \pm 0.17) \times 10^{-5}$ cm$^{-2}$ 
s$^{-1}$ sr$^{-1}$, where the error is systematics dominated. 
Our EGB spectrum is featureless, less intense, and softer than that 
derived from EGRET data.
\end{abstract}

\pacs{95.30.Cq,95.55.Ka,95.85.Pw,96.50.sb,98.70.Sa}
\keywords{} 
\maketitle

{\it Introduction:}
The high-energy diffuse \gray{} emission is dominated by \gray{s} produced
by cosmic rays~(CR) interacting with the Galactic 
interstellar gas and radiation fields, the so-called diffuse Galactic emission 
(DGE).
A much fainter component, commonly designated as ``extragalactic \gray{} 
background''~(EGB), was first detected against the bright 
DGE foreground by the {\it SAS-2} satellite \cite{Fichtel:1978} and 
later confirmed by analysis of the EGRET data \cite{Sreekumar:1998}. 
The EGB by definition has an isotropic sky distribution and is considered
by many to be the superposition of contributions from unresolved extragalactic 
sources including active galactic nuclei, starburst galaxies and \gray{} bursts (\cite{Dermer:2007} and
references therein) and truly-diffuse emission processes.
These diffuse processes include the possible signatures of large-scale 
structure formation \cite{Waxman:2000}, emission produced by the 
interactions of 
ultra-high-energy CRs with relic photons \cite{Kalashev:2009}, the 
annihilation or decay
of dark matter, and many other processes 
(e.g.,~\cite{Dermer:2007} and references therein).
However, the
diffuse \gray{} emission from inverse Compton (IC) scattering
by an extended Galactic halo of CR electrons 
could also be attributed to such a 
component if the 
size of the halo is large enough (i.e., $\sim 25$ kpc) \cite{Waxman:2004}.
In addition, \gray{} emission from CRs interacting in populations of 
small solar system bodies \cite{Moskalenko:2009} and the all-sky contribution
of IC scattering of solar photons with local CRs
can provide 
contributions \cite{Moskalenko:solarIC,Moskalenko:solarICerratum,Orlando:solarIC}.
Hence, an extragalactic origin for such a component is not clear, even though
we will use the abbreviation `EGB' throughout this paper.

In this paper, we present analysis and first results for the EGB derived 
from the \fermi\ Large Area Telescope (LAT) \cite{Atwood:2009} data.
Our analysis uses data from
the initial 10 months of the science phase of the mission.
Essential to this study is an event-level data selection with a higher 
level of background rejection than the standard LAT data selections, and 
improvements to the instrument simulation.
These have been made following extensive on-orbit studies of the 
LAT performance and of charged particle backgrounds.
Together, these improvements over the pre-launch modelling and background
rejection allow a robust derivation of the spectrum of the EGB that is 
not possible
with the standard low-background event selection.

{\it Data selection:}
The LAT is a pair-conversion telescope with a precision tracker and 
segmented calorimeter, each consisting of a $4\times 4$ array of 
16 modules, a segmented 
anti-coincidence detector (ACD) that covers the tracker array, 
and a programmable trigger and data acquisition system.
Details of the on-board and ground data processing
are given in \cite{Atwood:2009}.

The LAT ground processing makes use of the pre-launch 
background rejection scheme described in \cite{Atwood:2009}. 
The standard low-background event selection 
resulting from this multivariate analysis, 
termed ``diffuse'' class, has a Monte Carlo predicted 
background rate of $\sim 0.1$~Hz when integrated over the full instrument
acceptance $> 100$~MeV.
On-orbit investigations of the residual background of misclassified particles 
in the diffuse event selection indicated a higher level than predicted 
from pre-launch modelling. 
To reduce the residual particle background further, 
we developed an event selection comprised of the following four 
criteria in addition to the standard diffuse event classification: 
1) events are required to have a multivariate-analysis assigned \gray{} 
probability that is higher than the standard diffuse selection, with the 
required probability an increasing function with energy instead of a 
constant value as for diffuse class events;
2) the distance of extrapolated reconstructed particle tracks from the 
corners of the ACD must be higher than a set minimum value to remove
particles that enter the LAT in a region where the ACD has a lower than average
efficiency;
3) the average charge deposit in the silicon layers of the tracker is required 
to be small;
4) the reconstructed transverse shower size of events in the calorimeter 
is within a size range expected for electromagnetic showers.
The first two criteria assist in reducing the overall level of CR background. 
The second two criteria
provide an additional veto against hadronic showers and heavy 
ions that leak through the standard diffuse event classification.
In addition to these analysis cuts  
the particle background modelling
has been updated to be 
closer to the observed on-orbit charged particle rates. 
Furthermore, the instrument simulation now takes into account 
pile-up and accidental coincidence
effects in the detector subsystems that were not considered in the definition
of the pre-launch instrument response functions (IRFs) \cite{Rando:2009}.

\begin{figure}[t]
\includegraphics[width=8.5cm]{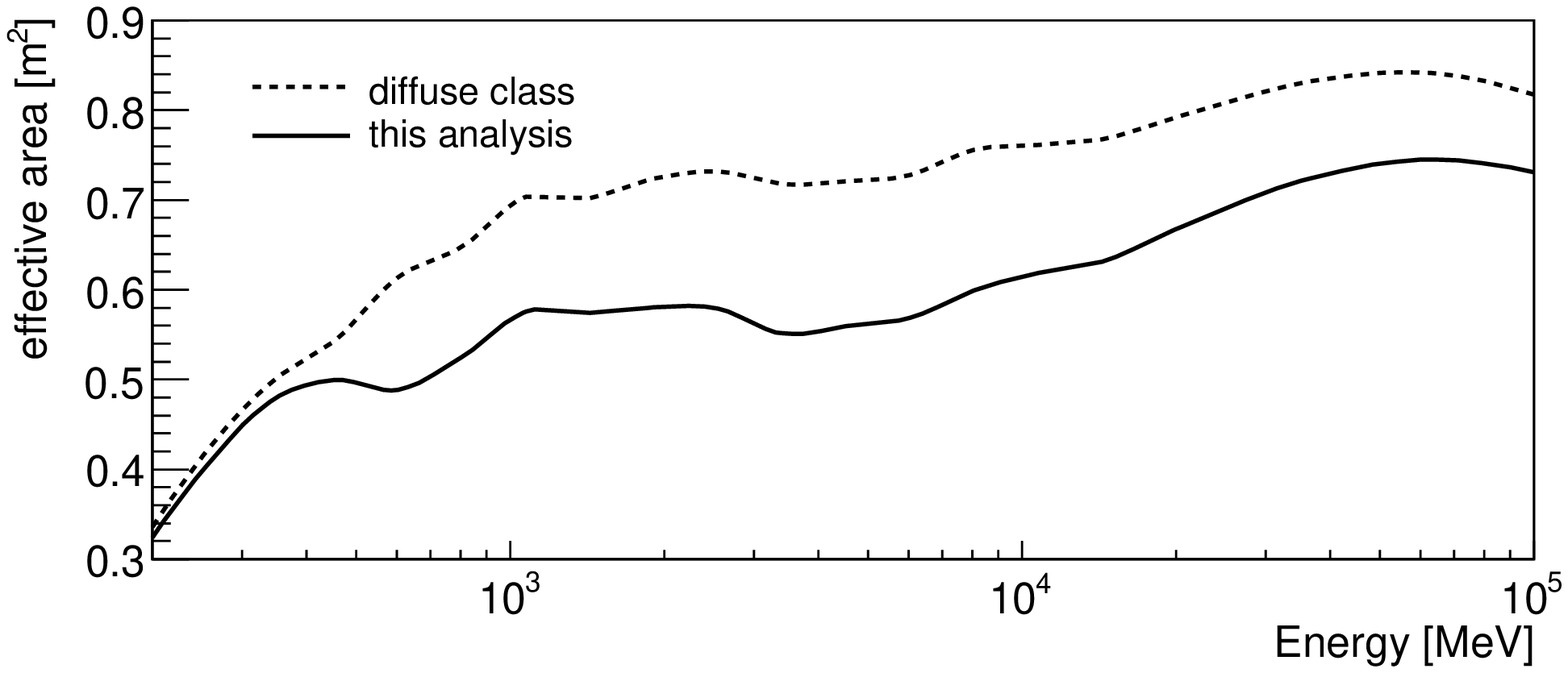}
\includegraphics[width=8.5cm]{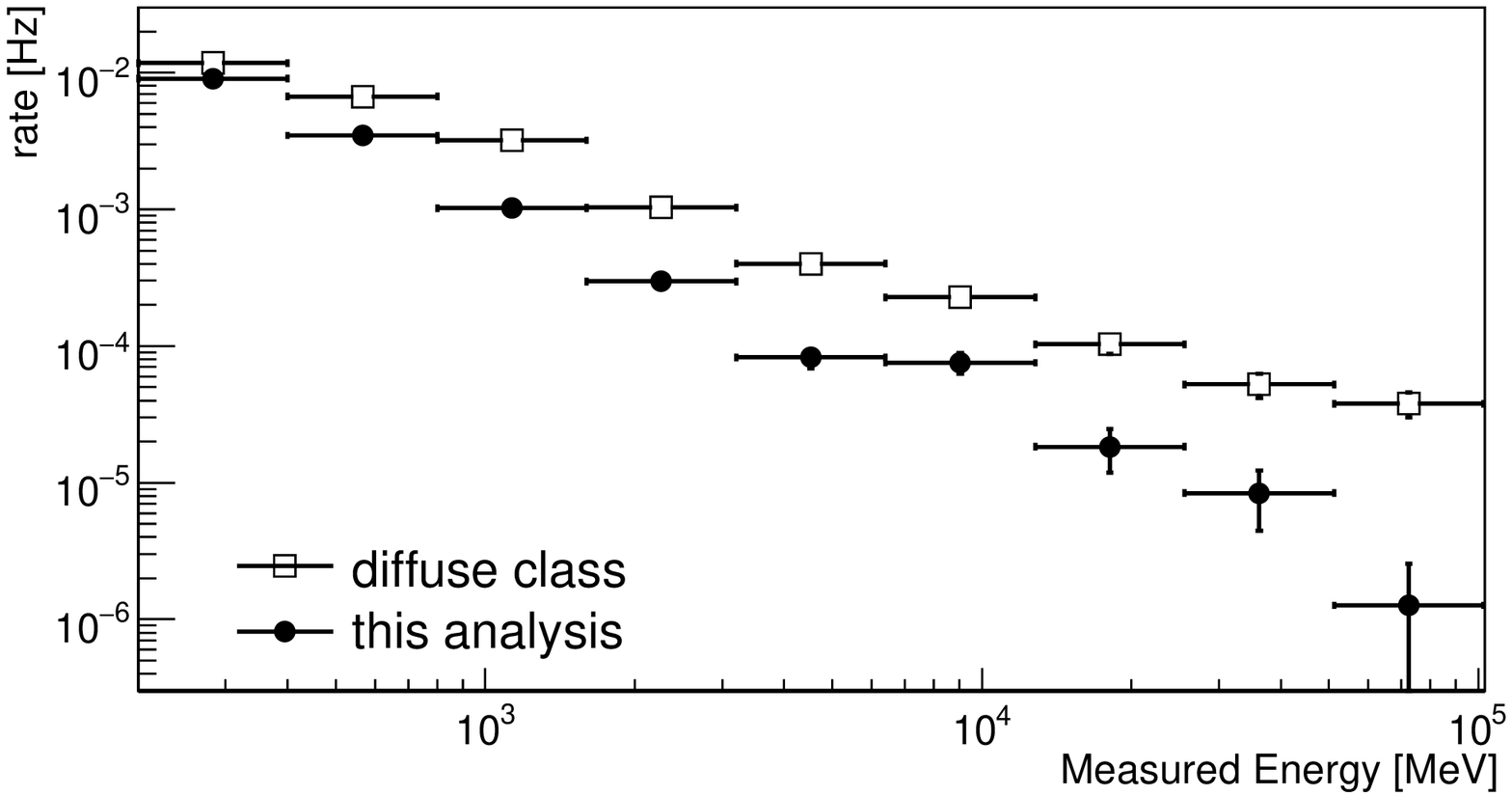}
\caption{\label{fig:effarea} Comparison of (a) LAT on-axis effective area and
(b) orbit-averaged CR background rate integrated over the FOV between the 
enhanced 
low-background event selection and the standard ``diffuse'' event selection.}
\end{figure}

\begin{figure}[t]
\includegraphics[width=8.5cm]{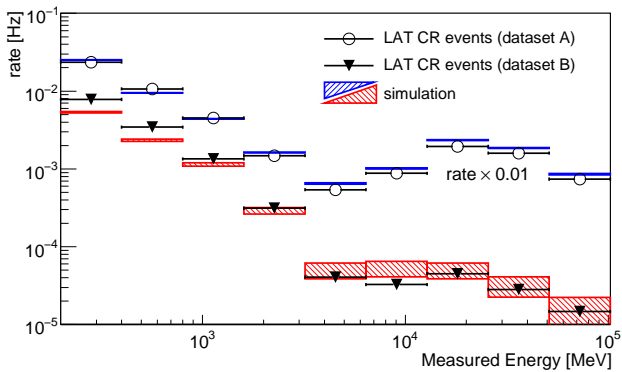}
\caption{\label{fig:residualbackground}Comparison of expected and 
measured orbit-averaged event rates for two CR dominated data samples.} 
\end{figure}

Figure~\ref{fig:effarea}a shows the on-axis effective area (${\rm A_{eff}}$) for
our enhanced low-background and standard diffuse selections, respectively.
The ${\rm A_{eff}}$ for the enhanced selection is reduced for 
energies $> 300$~MeV with a peak value $\sim 0.74$~m$^2$ 
compared to $\sim 0.84$~m$^2$ for diffuse class events.
The ${\rm A_{eff}}$ systematic uncertainties for our enhanced low-background
selection are of the same magnitude as those for the diffuse class events, 
evaluated by comparing the efficiencies of analysis cuts for data and 
simulation of observations of Vela: 10\% below 100 MeV, 
decreasing to 5\% at 560 MeV, and increasing to 20\% at 10 GeV and above.  
Figure~\ref{fig:effarea}b shows the orbit-averaged 
residual background rate of our enhanced low-background and standard 
diffuse selection, respectively, determined from our improved simulation.
With our enhanced event selection, the predicted background rejection is 
improved by a factor $1.3$--$10$.

We estimate the uncertainty of the residual CR background in our simulation
by comparing two CR background dominated LAT data samples to the predictions. 
The first data sample (A) contains all events passing the on-board filters, 
corresponding to a minimal background rejection level. 
The second data sample (B) contains events that pass the ``source'' 
classification in the standard analysis \cite{Atwood:2009} 
but fail to pass the more stringent diffuse selection.
This second sample corresponds to a very high level of background rejection but
is still dominated by charged particles compared to the standard 
diffuse selection, particularly at intermediate and high 
Galactic latitudes.
These particles are from the extreme tails of the CR distributions 
that are difficult to reject.
To both samples we apply the selection cuts 3) and 4), described above,
to remove the heavy ion and hadronic shower backgrounds 
observed on-orbit that are not modeled with sufficient accuracy in the 
simulation.
In addition,
to reduce the \gray{} fraction in both samples only events 
from Galactic 
latitudes $|b| \geq 45^\circ$ are used. 
The remaining \gray{} ``contamination'' is negligible for sample A due to 
the overwhelming CR rate. 
For sample B, the contamination 
is less than 10\% above 1~GeV, but is almost 30\% at 200 MeV, even at high 
Galactic latitudes.
To remove this \gray{} contamination from sample B 
we use the intensity maps from our 
fits to the data (see below), combined with the IRFs corresponding to 
sample B, to determine the expected \gray{} rate.
This is subtracted from the observed rate of sample B events.

Figure~\ref{fig:residualbackground} compares the 
orbit-averaged event rates measured by the LAT and predicted by our 
simulation 
for datasets A and B. 
At the minimal background rejection level represented by sample A
we find agreement within $\pm$~20\%.
This shows that the bulk of the remaining CR background is 
well described by the simulation 
after removing 
the particular class of heavy ion and hadronic shower events mentioned above.
For sample B, the 
agreement is within $+50$\%/$-30$\%, indicating the 
uncertainty in the description of the extreme tails of the CR 
distributions.
As these tails are responsible for the limiting background in 
the present analysis, we adopt the results for 
sample B as representative of the 
uncertainty on the expected residual CR background.

{\it Analysis:}
We use data taken in the nominal ``scanning'' mode from the 
commencement of scientific operations in 
mid-August 2008 to mid-June 2009. 
The data were prepared using the LAT Science Tools 
package, 
which is available from the \fermi{} Science Support 
Center\footnote{http://fermi.gsfc.nasa.gov/ssc/}.
Events satisfying our enhanced low-background event selection, coming from 
zenith angles $< 100^\circ$ (to greatly reduce the contribution by 
Earth albedo \gray{s}) and incidence angles within $65^\circ$ 
of the LAT z-axis (the LAT field-of-view) were used. 
This leaves 19 Ms of total observation time in the data set.
The energy-dependent exposure was calculated using the IRFs 
corresponding to our enhanced low-background event selection described above.

The photon counts and exposure were further processed using the 
\gadget\ package, part of a suite of tools we have developed to 
analyse the DGE \cite{Ackermann:2009}. 
Gamma-ray skymaps were generated using a HEALPix \cite{Gorski:2005} 
isopixelisation scheme at order 6 with 9 independent energy bins 
from 200 MeV to 102 GeV with \gadget\ used to simultaneously fit a DGE model, 
solar \gray{} emission, and sources (described below) to the resulting skymaps.
We only consider the Galactic latitude range $|b|>10^{\circ}$ in this analysis
where the DGE is more than an order of magnitude weaker than in the 
Galactic plane.

The model used for the large-scale DGE is based on the GALPROP 
code\footnote{http://galprop.stanford.edu, model id 77XvMM7A.}.
Recent improvements include use of the formalism and 
corresponding code for pion production in $pp$-interactions by
\cite{Kamae:2006,Kelner:2006}, 
a complete recalculation of the ISRF \cite{Porter:2008}, updated \hi{} 
and \htwo{} gas maps, including corrections to the total gas column density
derived from dust reddening maps \cite{Grenier:2005} 
an improved line-of-sight integration routine, 
and the addition of information from our ongoing studies of the 
DGE with the LAT \cite{LAT:HIEmissivity,LAT:GeVExcess}.
Cosmic-ray intensities and spectra are calculated
using a diffusive reacceleration CR transport model for a nominal 
halo size of 4 kpc, with a rigidity dependent diffusion 
coefficient that is consistent with available CR data for the B/C 
and $^{10}$Be/$^{9}$Be ratios, respectively.
We also consider bounding halo sizes 2 kpc and 10 kpc, with corresponding
self-consistently derived diffusion coefficients, 
since the size of
the CR halo is one of the principal uncertainties in the DGE foreground.
The injection spectra for CR protons and primary electrons are 
chosen to reproduce after propagation the locally measured spectra, including
the recently reported \fermi{} LAT CR electron spectrum \cite{Abdo:electrons}.
Gamma-ray emissivities are calculated using the propagated CR spectra and
intensities folded with the appropriate target distributions included in the
GALPROP code: \hi{}, \htwo{}, and \hii{} gas distributions for $\pi^0$-decay
and bremsstrahlung, and the ISRF for IC scattering.
Gamma-ray intensity skymaps are obtained by direct line-of-sight integration
of the calculated \gray{} emissivities.

For the dominant high latitude components, bremssstrahlung 
and $\pi^0$-decay emission from \hi{} and \hii{} in the local 
Galaxy ($7.5\,{\rm kpc} < R < 9.5 \, {\rm kpc}$) and IC emission, 
the intensities are fit to the LAT data via scale factors.
We use the GALPROP skymaps as templates with the 
component normalisations per energy bin 
as fit parameters. 
The sub-dominant high-latitude DGE components,
bremsstrahlung and  $\pi^0$-decay from \htwo{}, as
well as \hi{} and \hii{} outside the local region defined above, 
are taken from GALPROP predictions and do not vary in the fit.
All sources with test statistic above 200 
(i.e., larger than $\sim 14 \sigma$) found in the 
internal LAT 9-month source list are included with the flux per 
energy band per source as a fit parameter.
Weaker sources are included with fluxes derived from the LAT catalogue
analysis.
 In addition templates for the intensity of the \gray{} emission from CRs interacting
in the solar disk and radiation field \cite{Moskalenko:solarIC,Moskalenko:solarICerratum,Orlando:solarIC} that take into account the
relative exposure as the Sun transits the celestial sphere are included with 
their normalisations as fit parameters.

\begin{figure}[tb]
\includegraphics[width=8.5cm]{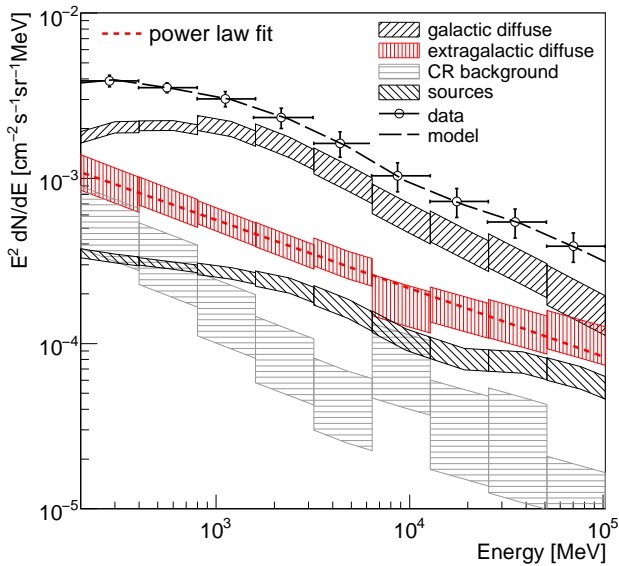}
\caption{\label{fig:data}LAT measured \gray{} intensity with fit results
for $|b| \geq 10^\circ$ including statistical and systematic errors.
Fit results by component are given in Table~\ref{tab:table1}.
Note LAT data are dominated by systematic uncertainties for the energy range
shown in the figure.}
\end{figure}

\begin{table*}[htb]
\caption{\label{tab:table1} Fit results and uncertainties for the EGB and
other components for $|b| \geq 10^\circ$.}
\footnotesize


\begin{tabular}{lccccccccc}
 & \multicolumn{9}{c}{Intensity integrated over energy band (cm$^{-2}$ s$^{-1}$ sr$^{-1}$)} \\
Energy in GeV & 0.2--0.4  &  0.4--0.8  &  0.8--1.6  &  1.6--3.2  &
                3.2--6.4  &  6.4--12.8  & 12.8--25.6  & 25.6--51.2  &
                51.2--102.4   \\	
Scale factor & \scale{6}   & \scale{7}   & \scale{7}   & \scale{8}   & \scale{8}   & \scale{9}   & \scale{9}   & \scale{9}   & \scale{10} \\
\hline\hline
 EGB & \err{2.4}{0.6} & \err{9.3}{1.8} & \err{3.5}{0.6} & \err{12.7}{2.1} & \err{5.0}{1.0} & \err{14.3}{4.0} & \err{6.3}{1.5} & \err{2.6}{0.7} & \err{11.1}{2.9} \\ 
\hline
Galactic diffuse (fit)    & \err{4.9}{0.4} & \err{25.9}{1.8} & \err{12.6}{1.3} & \err{50.7}{7.2} & \err{17.0}{3.0} & \err{50.0}{10} & \err{17.1}{3.6} & \err{6.1}{1.4} & \err{19.1}{5.2} \\
Galactic diffuse (model)  & 5.0 & 26.0 & 11.5 & 43.3 & 14.7 & 47.9 & 15.7 & 5.2 & 17.0 \\
~~~~IC (fit)            & \err{1.5}{0.1} & \err{6.8}{0.5} & \err{3.5}{0.4} & \err{16.1}{2.3} & \err{6.6}{1.2} & \err{23.3}{4.9} & \err{9.3}{2.1} & \err{3.9}{1.0} & \err{10.6}{3.7} \\
~~~~IC (model)          & 1.2 & 5.3 & 2.3 & 9.7 & 4.0 & 16.2 & 6.3 & 2.4 & 8.7 \\
~~~~local \hi{} (fit)   & \err{2.7}{0.2} & \err{15.4}{1.1} & \err{7.4}{0.8} & \err{28.3}{4.0} & \err{8.3}{1.5} & \err{20.6}{4.2} & \err{5.9}{1.2} & \err{1.6}{0.4} & \err{7.0}{2.2} \\
~~~~local \hi{} (model) & 3.1 & 17.0 & 7.6 & 27.6 & 8.7 & 26.0 & 7.7 & 2.3 & 6.8 \\
Sources               & \err{0.8}{0.1} & \err{3.8}{0.2} & \err{1.7}{0.1} & \err{7.2}{0.8} & \err{2.7}{0.4} & \err{9.0}{1.3} & \err{3.4}{0.5} & \err{1.5}{0.2} & \err{6.3}{1.0} \\
CR background         & \err{1.4}{0.6} & \err{4.2}{1.7} & \err{1.0}{0.4} & \err{2.8}{1.2} & \err{0.8}{0.4} & \err{6.3}{3.0} & \err{1.4}{0.8} & \err{0.6}{0.4} & \err{0.9}{0.9} \\
Solar         & \err{0.1}{0.01} & \err{0.4}{0.04} & \err{0.2}{0.02} & \err{1.0}{0.2} & \err{0.4}{0.2} & \err{1.7}{0.4} & \err{0.7}{1.6} & \err{0.1}{0.04} & \err{0.8}{0.5} \\
\hline
LAT       & \err{9.6}{0.8} & \err{44.0}{3.0} & \err{18.8}{2.0} & \err{72.9}{10} & \err{25.3}{4.5} & \err{81.3}{16} & \err{28.3}{5.7} & \err{10.6}{2.1} & \err{37.9}{7.7} \\
\hline
 & & & & & & & & & \\
 & \multicolumn{9}{c}{Foreground modeling related uncertainty in cm$^{-2}$ s$^{-1}$ sr$^{-1}$} \\
\hline\hline 

\hi{} column density    & $+0.1$/$-0.3$ & $+0.1$/$-1.7$ & $+0.1$/$-0.9$ & $+0.1$/$-3.6$ & $+0.1$/$-1.1$ & $+0.1$/$-2.4$ & $+0.1$/$-0.9$ & $+0.1$/$-0.2$ & $+0.1$/$-1.1$ \\
IC + halo size      & $+0.1$/$-0.2$ & $+0.1$/$-0.8$ & $+0.1$/$-0.5$ & $+0.1$/$-1.8$ & $+0.1$/$-0.5$ & $+0.1$/$-0.7$ & $+0.3$/$-0.3$ & $+0.4$/$-0.1$ & $+2.9$/$-0.5$ \\
CR propagation model   & $+0.1$/$-0.3$ & $+0.1$/$-1.1$ & $+0.1$/$-0.6$ & $+0.1$/$-0.8$ & $+0.1$/$-0.3$ & $+0.1$/$-1.2$ & $+1.4$/$-0.1$ & $+0.4$/$-0.1$ & $+3.0$/$-0.1$ \\
Subregions of $|b|>10^{\circ}$ sky  & $+0.2$/$-0.3$ & $+0.8$/$-1.5$ & $+0.4$/$-0.9$ & $+1.9$/$-2.1$ & $+0.7$/$-0.5$ & $+2.5$/$-1.9$ & $+1.0$/$-1.5$ & $+0.5$/$-0.3$ & $+2.7$/$-0.9$ \\
\hline
\end{tabular}



\end{table*}

{\it Results:}
Figure~\ref{fig:data} shows the \gray{} intensity measured by the LAT and 
the fit results for the Galactic latitude range $|b| \geq 10^\circ$. 
Table~\ref{tab:table1} summarises the numerical values and uncertainties,
including the intensity values for the individually fitted DGE components 
that are not distinguished in figure~\ref{fig:data} for clarity.
The residual intensity obtained after fitting the DGE model components, solar
emission, and sources is the sum of 
CR background and EGB. The simulation is used to estimate the CR background and uncertainty, 
as described earlier. The CR background is isotropic when averaged over the data taking period
in this paper and is subtracted to obtain the EGB intensity. Additional figures for different 
latitude bands and regions of the sky can be found online \cite{epaps}.

Our formal uncertainty on the EGB comes from the fit using the nominal model.
However, the RMS of the residual count fraction between LAT data and our model 
for energies above 200 MeV is 8.2\%, when averaged over regions of $13.4$ deg$^2$ to ensure sufficient 
statistics.
This is larger than the 3.3\% value expected solely from statistical 
fluctuations. We also see correlation of the residual count 
fraction with structures in the DGE model skymaps. 
This suggests a limitation in the accuracy of the description of the 
DGE model. 
We investigated the uncertainty on the EGB flux related to the
DGE components by varying the relevant parameters in the 
model and re-evaluating the fits for $|b| > 10^\circ$.
At high latitudes, the model parameters principally affecting the DGE are:
the change of the IC emission with 
different halo sizes and the calculation of the IC emission using the 
anisotropic/isotropic formalism \cite{MS:2000} 
(IC + halo in Table~\ref{tab:table1}), variations of 
the CR source distribution and \Xco{} gradient (CR propagation model), 
and how assumptions used to derive 
\hi{} column densities from radio data and dust reddening 
measurements affect the distribution of \hi{} in the local region (\hi{} column density).
To quantify the uncertainty connected to the 
residual count fraction, we used the nominal model and examined the 
variation of the derived EGB 
when different subregions of the $|b| > 10^\circ$ sky are 
fitted (Subregions of $|b|>10^{\circ}$ sky).
No single component dominates the uncertainties shown in the lower 
half of Table~\ref{tab:table1}. 
We caution that the uncertainties 
for the model components cannot be assumed to be independent.
Hence, there is no simple relationship between the combination
of individual components and the total formal uncertainty. 

The large statistics allow sub-samples of the total data set to be used as
a cross check.
We repeated our analysis for events passing our enhanced selection with
1) different onboard trigger 
rates and 2) conversions in the thin or thick sections of the 
tracker \cite{Atwood:2009}.
The first sub-sample ensures that we have properly estimated the residual CR 
background, while the second checks that
the small fraction of misreconstructed Earth albedo events that enter the LAT in the back section 
do not affect the result.
The derived EGB spectrum for these sub-samples is completely 
consistent with that derived from the full data set using the same analysis
procedure.

\begin{figure}[tb]
\includegraphics[width=8.5cm]{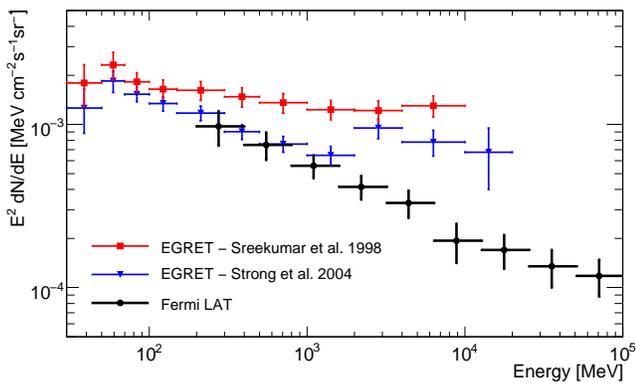}
\caption{\label{fig:sed} EGB intensity derived in this work compared with
EGRET-derived intensities taken from table 1 in \cite{Sreekumar:1998} and 
table 3 in \cite{Strong:2004}.
Our derived spectrum is compatible with a simple power-law with index 
$\gamma = 2.41 \pm 0.05$ and intensity 
$I(>100\,{\rm MeV}) = (1.03 \pm 0.17) \times 10^{-5}$~cm$^{-2}$~s$^{-1}$~sr$^{-1}$ where the uncertainties are systematics dominated.}
\end{figure}

Finally, we note that our analysis also indicates a significant detection 
of the combined solar disk and extended solar IC emission.
This finding will be explored in more detail in a separate study.

{\it Discussion:}
Figure~\ref{fig:sed} shows the spectrum of the EGB above 200 MeV derived
in the present analysis, and  
from EGRET data \cite{Sreekumar:1998,Strong:2004}.
Our intensity extrapolated to 100 MeV based on the power-law fit, 
$I(> 100 \, {\rm MeV}) = (1.03 \pm 0.17) \times 10^{-5}$~cm$^{-2}$~s$^{-1}$~sr$^{-1}$, is significantly lower than that obtained from EGRET data: 
$I_{\rm EGRET} (> 100\, {\rm MeV}) = (1.45 \pm 0.05) \times 10^{-5}$ cm$^{-2}$ s$^{-1}$ sr$^{-1}$ \cite{Sreekumar:1998}.
Furthermore, our spectrum is compatible with a featureless power law with
index $\gamma = 2.41\pm0.05$.
This is significantly softer than the EGRET spectrum with 
index $\gamma_{\rm EGRET} = 2.13\pm0.03$ \cite{Sreekumar:1998}.
To check that the different spectra are not due to the instrumental 
point-source sensitivities, we adopt 
$F(> 100 \, {\rm MeV}) = 10^{-7}$~cm$^{-2}$~s$^{-1}$, comparable to the 
average EGRET sensitivity, and attribute the flux of all detected LAT 
sources below this threshold to the EGB.
We obtain an intensity 
$I_{res}(> 100\, {\rm MeV}) = (1.19 \pm 0.18) \times 10^{-5}$~cm$^{-2}$~s$^{-1}$ sr$^{-1}$  
and a spectrum compatible with a power-law with index 
$\gamma_{res}  = 2.37 \pm 0.05$. 
Therefore, the discrepancy cannot be attributed to a lower threshold for 
resolving point sources. 
Our EGB intensity is comparable to that obtained in the EGRET 
re-analysis by \cite{Strong:2004} with an updated DGE 
model, $I_{\rm SMR} (> 100\, {\rm MeV}) = (1.11 \pm 0.1) \times 10^{-5}$ cm$^{-2}$ s$^{-1}$ sr$^{-1}$.
However, our EGB spectrum does not show the distinctive harder spectrum
above $\gtrsim 1$~GeV and peak at $\sim 3$~GeV found in the same 
EGRET reanalysis. We note that the LAT-measured spectra are softer above $\gtrsim 1$~GeV
than those measured by EGRET also for the DGE at intermediate latitudes 
\cite{LAT:GeVExcess} and for the Vela Pulsar \cite{Abdo:Vela}.

{\it Acknowledgements:}
The \fermi{}  LAT Collaboration acknowledges support from a number of agencies 
and institutes for both development and the operation of the LAT as well as 
scientific data analysis. These include NASA and DOE in the United States,
CEA/Irfu and IN2P3/CNRS in France, ASI and INFN in Italy, MEXT, KEK, 
and JAXA in Japan, and the K.~A.~Wallenberg Foundation, the Swedish 
Research Council and the National Space Board in Sweden. Additional 
support from INAF in Italy and CNES in France for science analysis 
during the operations phase is also gratefully acknowledged.
GALPROP development is partially funded via NASA grant NNX09AC15G.
Some of the results in this paper have been derived using the 
HEALPix \cite{Gorski:2005} package.

\bibliography{extragalactic}

\clearpage

\newpage

\section{The Spectrum of the Isotropic Diffuse Gamma-Ray Emission Derived From First-Year \fermi{} Large Area Telescope Data. 
Supplementary online material.}

\subsection{Abstract}
Supplementary material concerning the analysis presented in "The Spectrum of the Isotropic Diffuse Gamma-Ray Emission 
Derived From First-Year \fermi{} Large Area Telescope Data" is presented here.

\subsection{Galactic diffuse model}

\begin{figure}[htb]
\includegraphics[width=8.5cm]{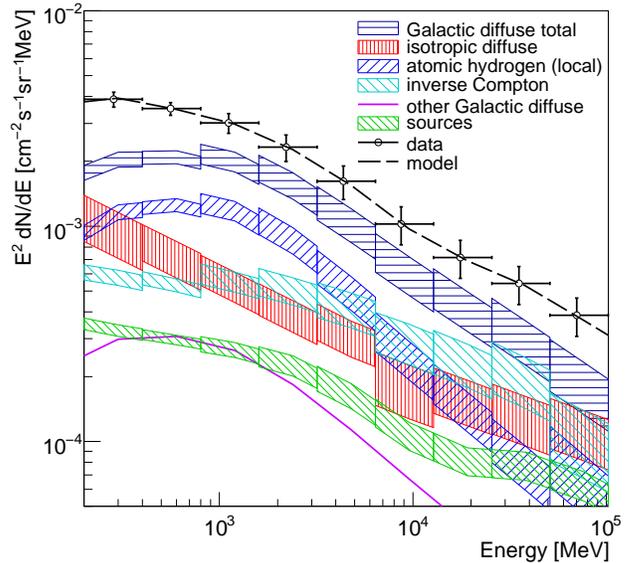}
\caption{\label{fig:galcomp}LAT measured \gray{} intensity with fit results
for $|b| \geq 10^\circ$ including statistical and systematic errors.
Independently fitted components of the galactic diffuse emission are shown individually.
``Other Galactic diffuse'' denotes the sum of the \htwo{} and non-local \hi{} and \hii{} 
contributions which are included using the model intensities (i.e., no fit).
Note LAT data are dominated by systematic uncertainties for the energy range
shown in the figure. Ths intensity of the solar emission considered in the model
is below the range shown in the figure. }
\end{figure}

Figure \ref{fig:galcomp} displays the individually fitted contributions
to the galactic diffuse emission arising from inverse Compton emission and 
interaction of CRs with atomic hydrogen in the local Galaxy (7.5 kpc $\leq$ r $\leq$ 9.5 kpc)
via bremsstrahlung and pion decay. 
These contributions are omitted in figure 3 of the published article for visual clarity, 
but numerically available in table 1. The intensity is averaged over galactic latitudes $|b| \geq 10^{\circ}$.
The cosmic-ray contamination is subtracted
from the EGB component shown here. Its contribution can be found in figure 3 of the published article.
The errors shown in the graphs are the quadratic sum of statistical and 
systematic errors due to uncertainties in the LAT effective area and CR background subtraction.

\subsection{Comparisons for different sky regions}

Figure 3 in the published article compares the modeled with the measured \gray{} intensity 
averaged over all galactic latitudes  $|b| \geq 10^{\circ}$. To illustrate the good agreement of the 
used \gray{} emission model, i.e. galactic diffuse emission, EGB, point sources and solar emission,  
over the whole fitted region, it is interesting to display this comparison for different sub-regions
of the sky.

\begin{figure*}[htb]
\includegraphics[width=8.5cm]{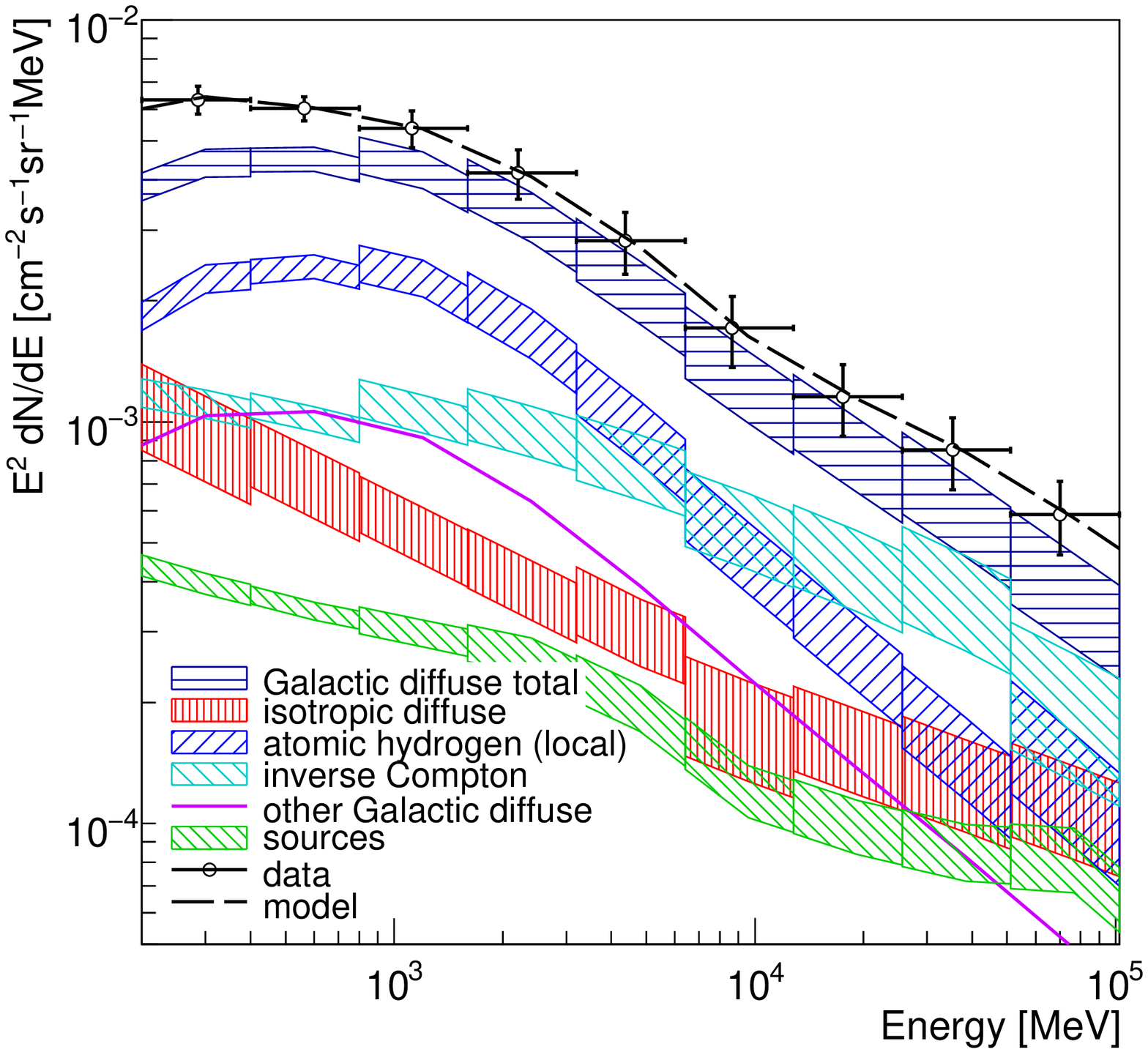}
\includegraphics[width=8.5cm]{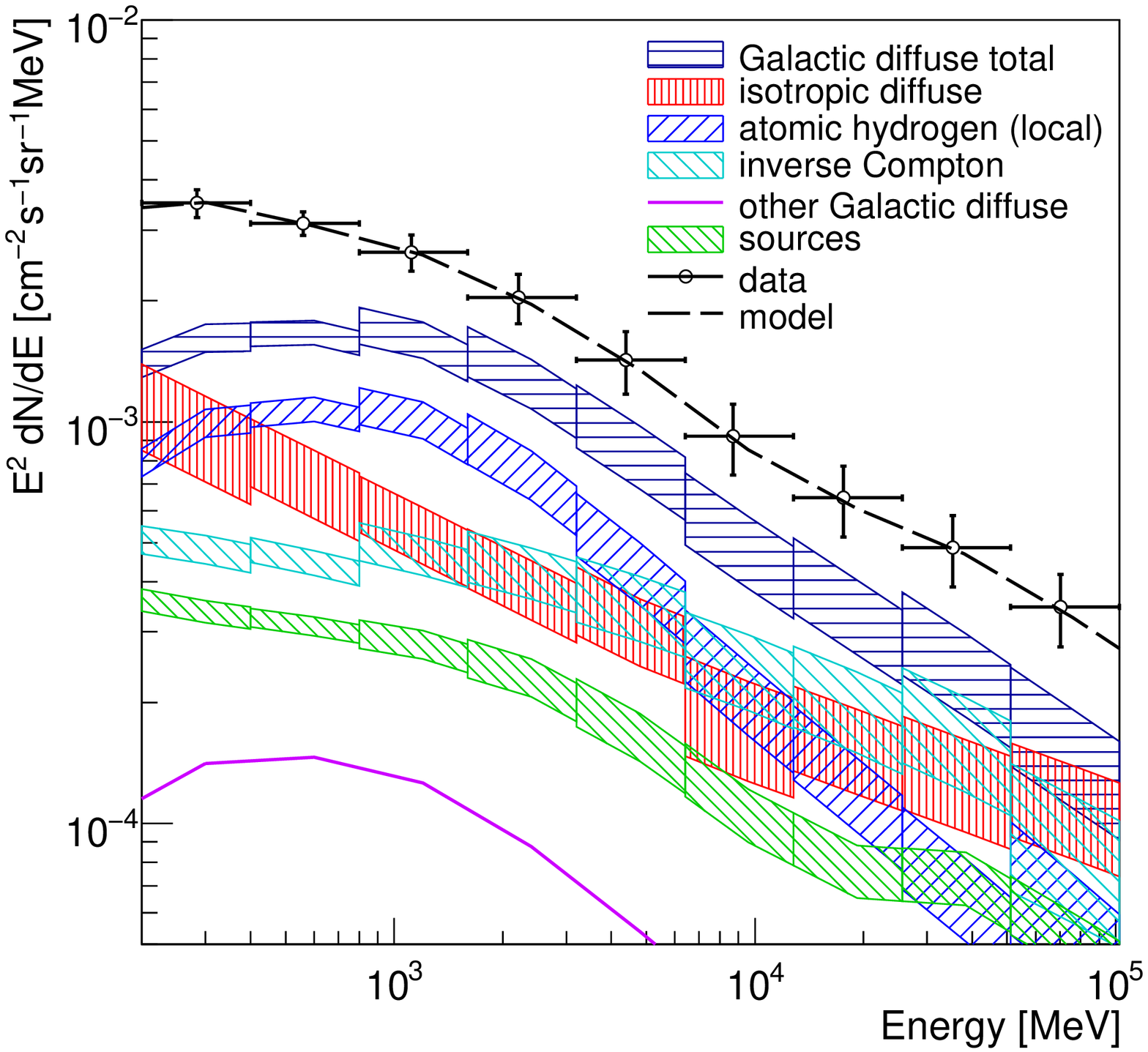} \\
\includegraphics[width=8.5cm]{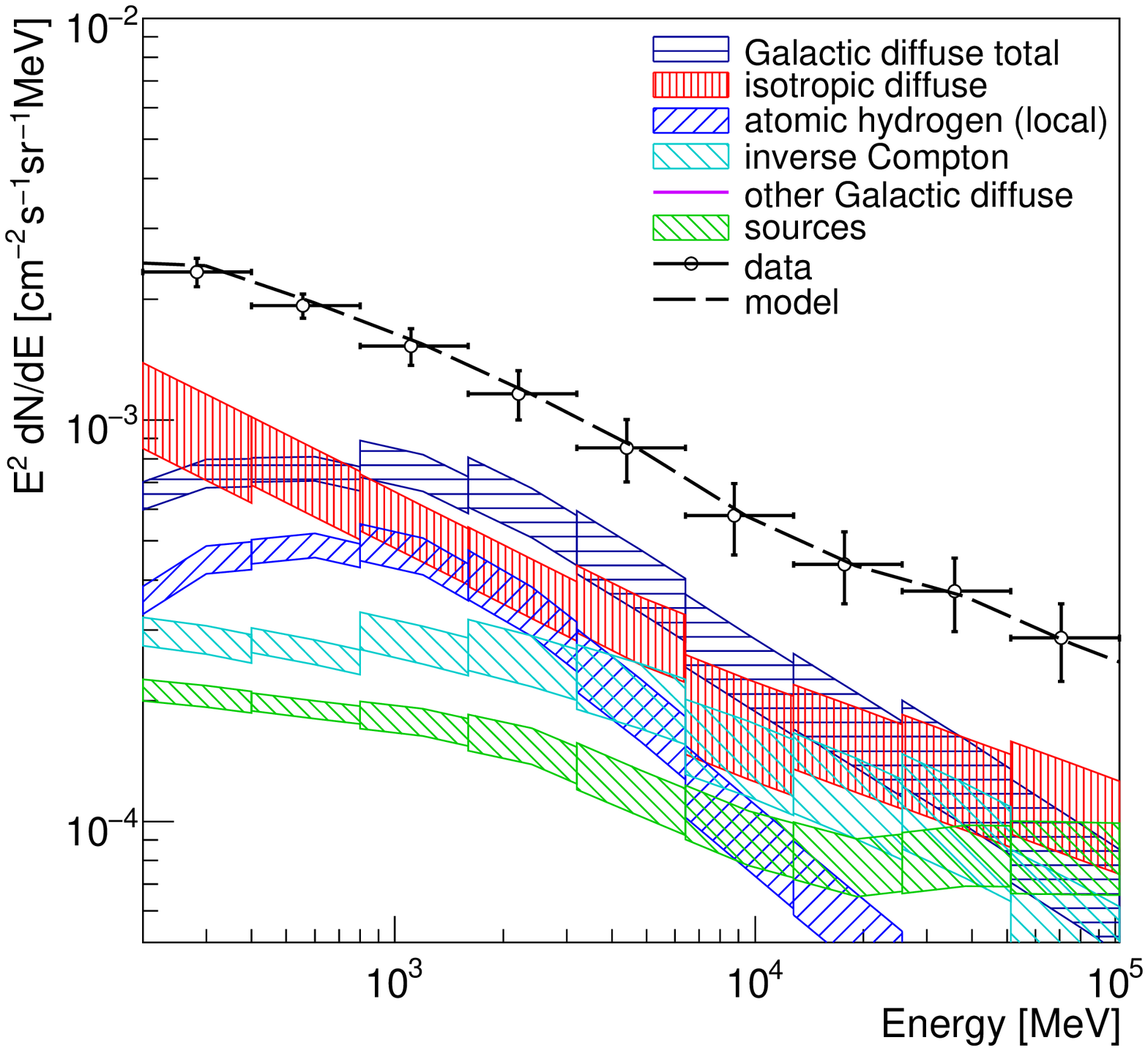}
\caption{\label{fig:latitudes} LAT measured intensity compared to 
the \gray{} emission model used in the derivation of the EGB
averaged over different ranges in Galactic latitude. The regions shown are 
$10^{\circ} \leq |b|  \leq 20^{\circ}$ (upper left), $20^{\circ} \leq |b| \leq 60^{\circ}$ (upper right) 
and $|b| \geq 60^{\circ}$ (lower center). ``Other Galactic diffuse'' denotes the sum of the \htwo{} and non-local \hi{} and \hii{} 
contributions which are included using the model intensities (i.e., no fit).
Errors include statistical and systematic errors. LAT data are dominated by systematic uncertainties for the energy range
displayed in the figure.}
\end{figure*}

\begin{figure*}[htb]
\includegraphics[width=8.5cm]{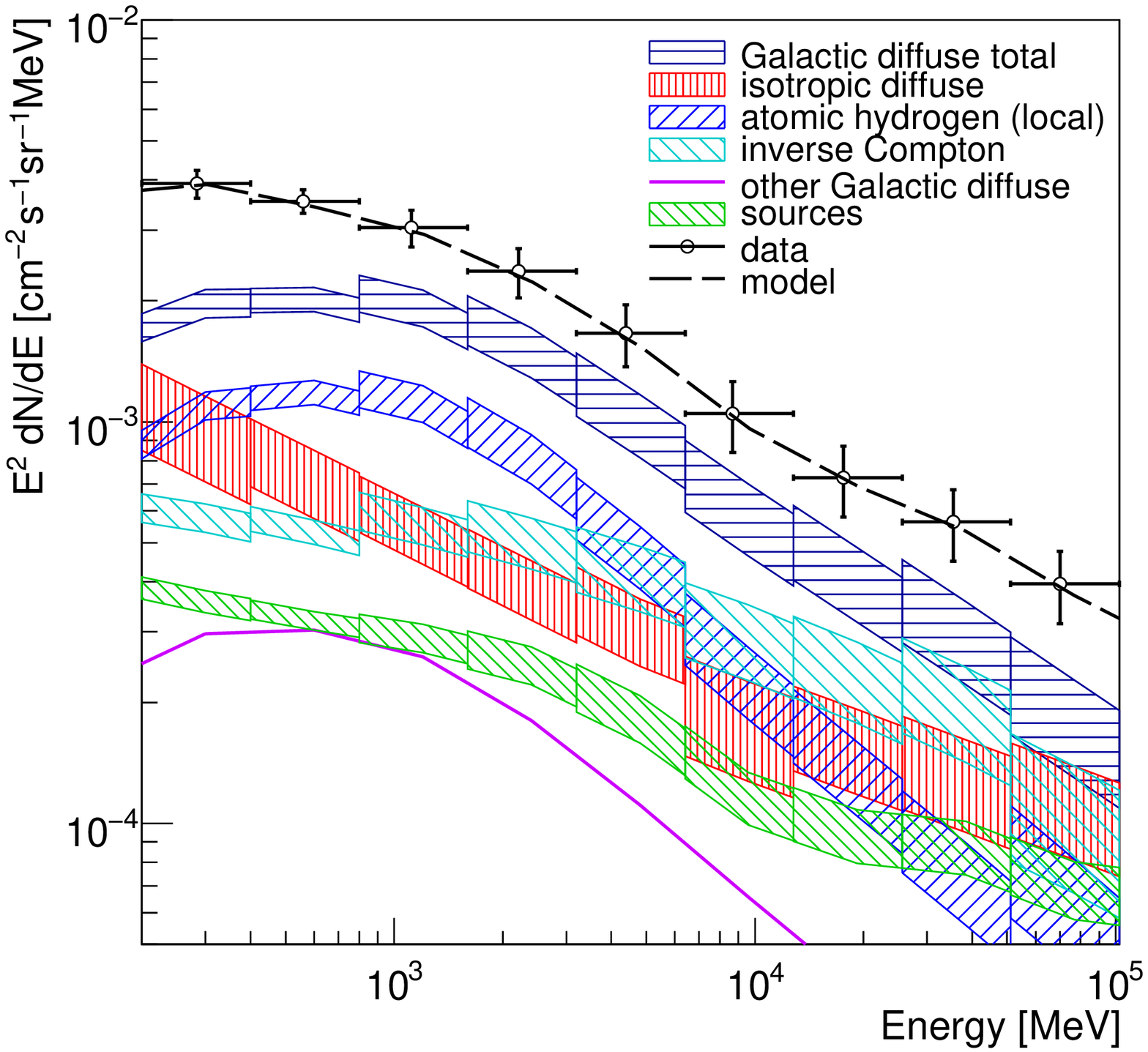}
\includegraphics[width=8.5cm]{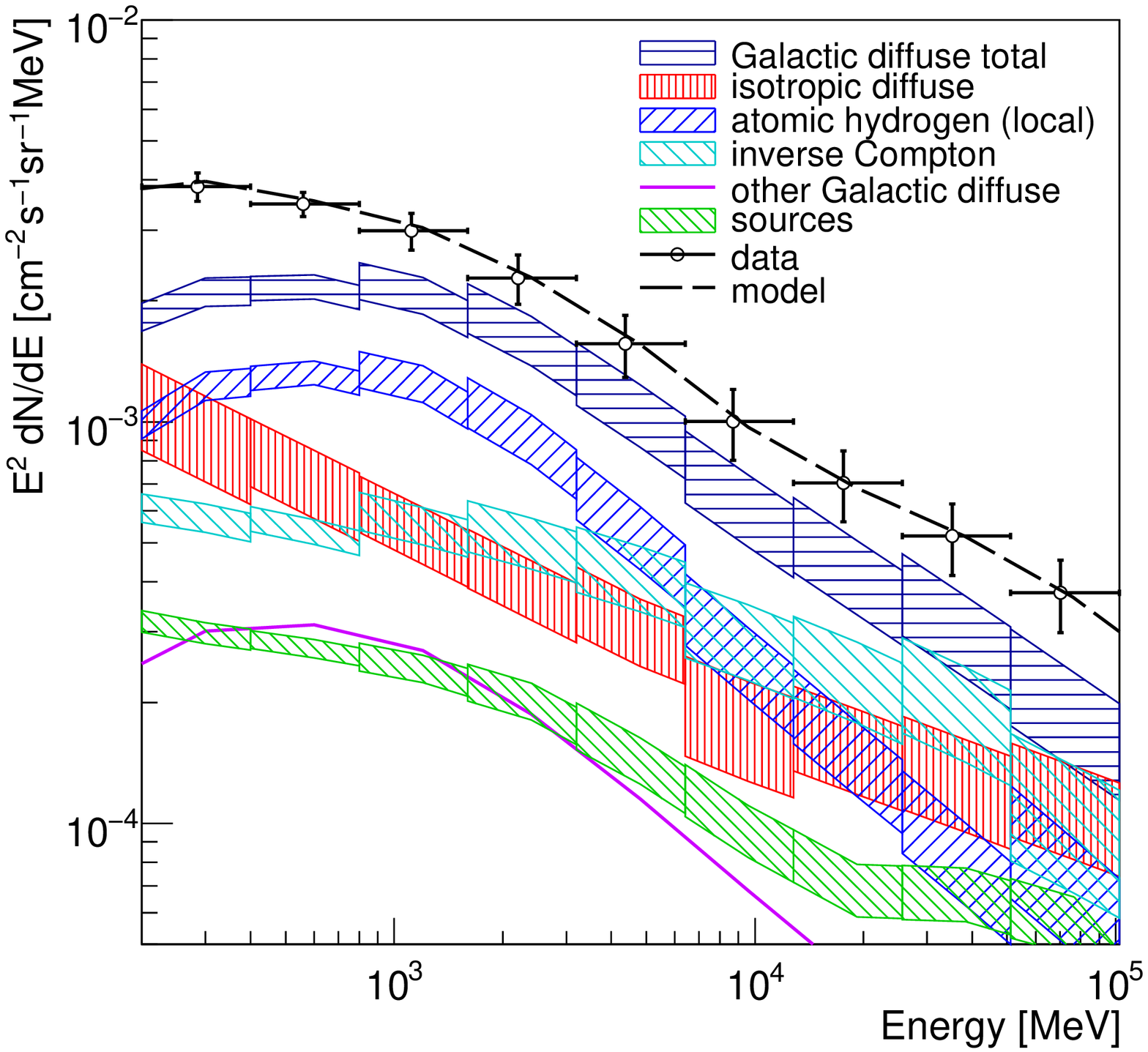} \\
\includegraphics[width=8.5cm]{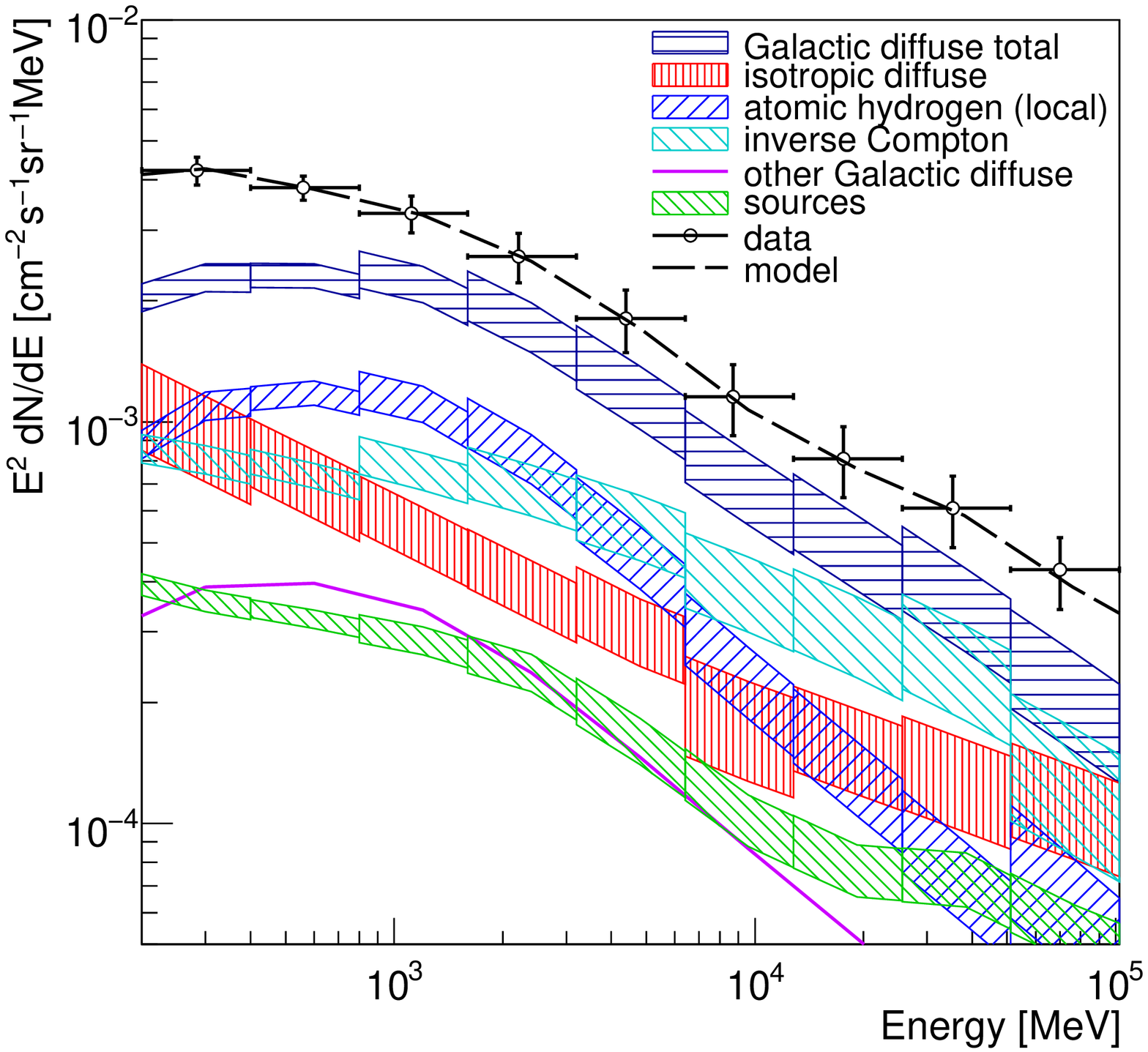}
\includegraphics[width=8.5cm]{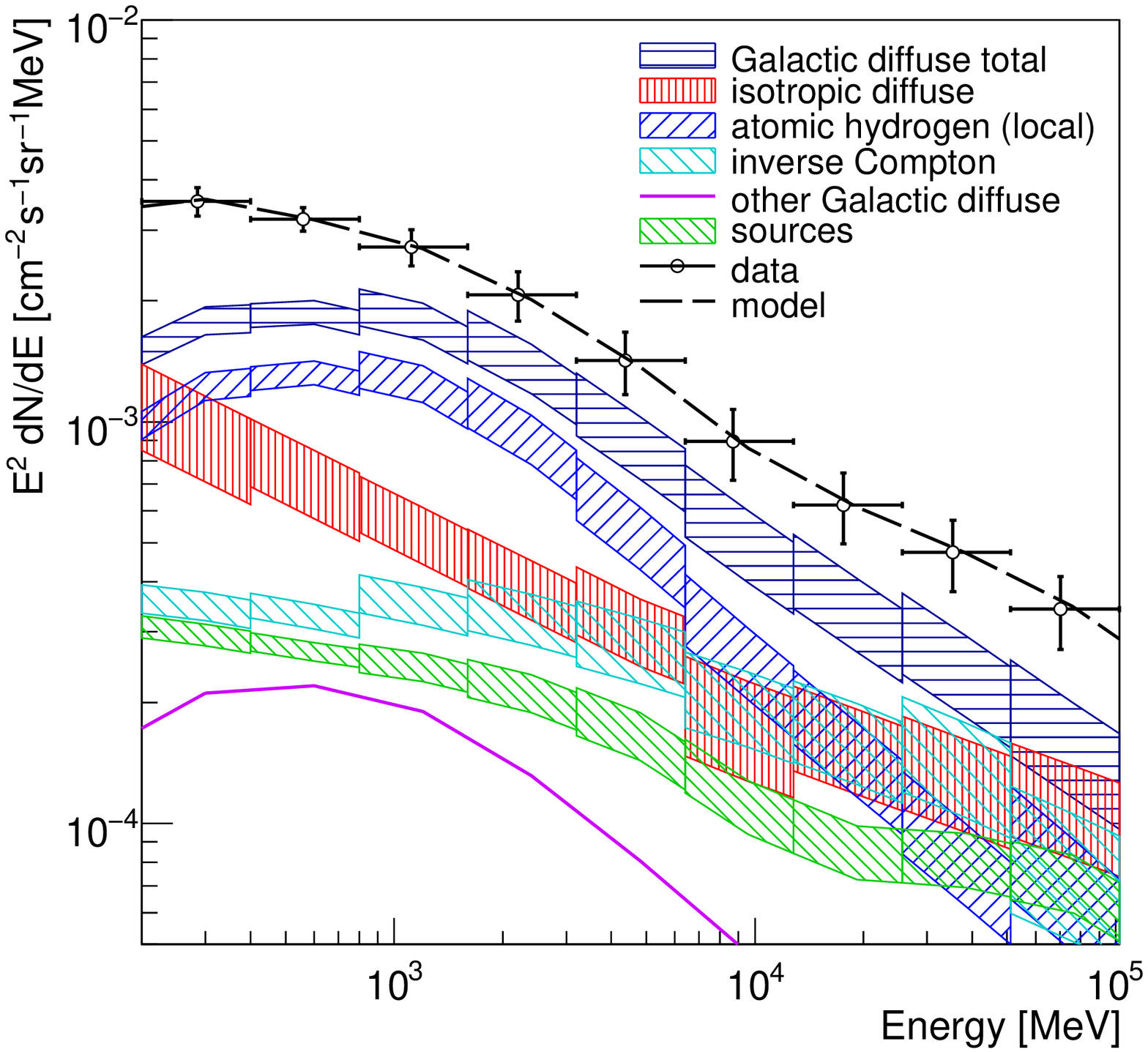}

\caption{\label{fig:hemispheres} LAT measured intensity compared to 
the \gray{} emission model used in the derivation of the EGB
averaged over different hemispheres on the sky for Galactic latitudes  $|b| \geq 10^{\circ}$. The 
hemispheres shown are centered at the North Galactic pole (upper left), the South Galactic pole (upper right), 
the Galactic center (lower left) and anti-center (lower right).
``Other Galactic diffuse'' denotes the sum of the \htwo{} and non-local \hi{} and \hii{} 
contributions which are included using the model intensities (i.e., no fit).
Errors include statistical and systematic errors. LAT data are dominated by systematic uncertainties for the energy range
displayed in the figure.}
\end{figure*}

Figure \ref{fig:latitudes} shows the intensities of different components in the model when averaged 
over independent Galactic latitude ranges covering low, mid and high galactic latitudes,
$10^{\circ} \leq |b|  \leq 20^{\circ}$, $20^{\circ} \leq |b| \leq 60^{\circ}$ and $|b| \geq 60^{\circ}$.
Figure \ref{fig:hemispheres} shows the intensities of different components in the model when averaged 
over different hemispheres. The hemispheres considered are centered at the North Galactic pole ($b \geq 0^{\circ}$),
the South Galactic pole ($b \leq 0^{\circ}$), the Galactic center ($270^{\circ}\leq l \leq 90^{\circ}$) 
and anticenter ($90^{\circ}\leq l \leq 270^{\circ}$).
 Furthermore, Galactic latitudes $|b|<10^{\circ}$ are excluded from all hemispheres.

We emphasize that the intensities of the components shown in the figures are from the 
single gamma-ray emission model used in the analysis. In particular, the EGB component 
in each figure is identical because it is isotropic by construction 
and thus its average intensity does not vary across the sky.
The accuracy of the model with respect to the different sub-regions of the sky shown can be judged by comparing 
the total predicted model intensity (black line in figure \ref{fig:latitudes}) to the measured 
gamma-ray intensities per energy band represented by the data points. 

\end{document}